\documentclass[aps,prd,superscriptaddress,showpacs,nofootinbib,preprintnumbers,a4]{revtex4}
\usepackage{amsmath}
\usepackage{amsfonts}
\usepackage{graphicx}
\usepackage{dcolumn}
\usepackage{hyperref}
\usepackage{amssymb}

\begin{document}

\title{A Unified View of the Basic Forces}
\bigskip

\author{Naresh Dadhich {\footnote{e-mail address:{nkd@iucaa.ernet.in}}}}
\affiliation{IUCAA, Post Bag 4, Ganeshkhind, Pune 411 007, INDIA}


\begin{abstract}
In this essay we wish to seek a unifying thread between the basic forces. We 
propose that there exists a universal force which is shared by all that physically exists. Universality is characterized by the two properties: (i) universal linkage and (ii) long range. They uniquely identify Einstein gravity as the unversal force. All other forces then arise as these properties are peeled off. For instance, relaxing (i) but retaining (ii) will lead to Maxwell electromagnetic force. This unified outlook makes interesting suggestions and predictions: if there exists a new force, it can only be a short range non-abelian vector or a scalar field, and there should exist in an appropriate space duality relations between weak and electric, and between strong and gravity. 
\end{abstract}
\maketitle

We propose that all that physically exists in the Universe interacts with 
each-other through a universal force. Universality is characterized by the following two properties: \\

(a) Universal Linkage. It links/interacts with all particles whether massive or massless. The universal parameter which is shared by all particles is energy, and since force also possesses energy, hence it also interacts with itself. \\

(b) Long Range. Since it has to interact with all, it must be present everywhere and hence it should be of long range. That means its propagator is massless.
For long range force, Gauss law should apply to give the conservation of 
charge which produces the force. \\

Clearly we have four possibilities: one, the universal force obeying both 
(a) and (b), second, not (a) but (b), third, (a) but not (b) and finally neither of the two. It will turn out that the universal force is uniquely Einstein 
gravity which is the mother force from which the other three emerge as we 
peel off the conditions one by one. The other three forces will respectively 
be electromagnetic, weak and strong forces. \\ 

Gravity is a second rank metric field represented by massless spin 2 particle 
and hence it can give on peeling off spin 1 and 
spin 0 fields. Spin 1 is a vector field which could be abelian or non-abelian 
with different symmetry properties. This offers a much richer space of 
possibilities. Abelian vector field gives Maxwell electromagnetic force while 
weak and strong forces are given by non-abelian field with $SU(2)$ and $SU(3)$  symmetry respectively. Of course we can have spin 0 scalar fields like Higgs 
scalar. \\

We shall in the following section discuss these four possibilities giving 
rise to four basic forces which will be followed by discussion of suggestions 
and predictions this perspective presents. \\

\section{Four Forces}
\subsection{Gravity}

Let us first consider the universal force satisfying the above two properties. Since the force interacts with both massive as well as massless particles, hence particle equation of motion under this force should be independent of 
mass. Like the motion under no force is a geometric property (straight line) 
of Euclidean space, so should be the case for motion under universal force. 
Because absence of force and force which acts on everything and present everywhere are both universal statements. Like the former is a property of flat spacetime, so should become the latter a property of curved spacetime~\cite{n1,n2,n3}. It turns that the universal force can only be described by curvature of spacetime and it can be nothing other than Einsteinian gravity, described by general relativity (GR). Einstein's equation for gravity follows from the Riemann curvature tensor through the Bianchi differential identity. Note that dynamics of 
universal force is self determined by the geometry of spacetime and it can not be prescribed. This is the distinguishing and unique feature of gravity - the universal force.   \\

This shows that the universal force obeying the above two properties is gravity and hence there can exist no other universal force. This is a very strong and general statement. Its universality has been further probed~\cite{n3,n4,n5,n6}. It is argued that like $2$ and $3$ 
dimensions are not big enough to accommodate propagation of gravity, similarly the usual $4$-dimensional spacetime is not big enough to fully accommodate self interaction of gravity~\cite{n6}. Gravity is a self interactive force and self interaction can only be evaluated by an iteration process. The first iteration is however included in Einstein equation through square of first derivative of the metric. It turns out that physical realization of second iteration can not be done in dimensions less than $5$. Hence, we have to go to higher dimensions for realization of second iteration of self interaction. This is a purely classical motivation for higher dimension for description of gravity.\\ 

Thus universal force is a tensor field described by a metric, $g_{ab}$, and its dynamics follows entirely from the spacetime curvature without prescription through the Bianchi differential identity satisfied by the Riemann curvature tensor~\cite{n1,n3}. It is a massless spin 2 tensor field. \\

\subsection{Electromagnetic Force}

We now relax the property (a) but retain (b), and so we have the case of no universal linkage but long range. It links to particles 
having some specific `charge' which distinguishes them from others. There are particles for which 
this charge is zero. That means the charge can be neutralized and hence it has to be bipolar, positive/negative. Since the force is long range, its propagation will however be massless. There 
should be conservation of charge. Then Gauss law will demand that force be 
inverse square law. Bipolarity of charge implies that the force is a vector 
field and the inverse 
square law will require the analogue of the Poisson equation to be a covariant 
$4$-vector equation with $4$-current as the source for the $4$-potential. 
This will indeed be Maxwell equation. The force in question could then be 
nothing other than Maxwell electromagnetic filed. \\

By relaxing the property (a) above we have arrived at a massless 
vector field. It has spin 1 and $U(1)$ gauge symmetry. It is thus  
an Abelian vector field. Like gravity, is this unique ? What we have shown above is that dynamics of 
force will be governed by Maxwell equation which follow from inverse square law and bipolarity of the charge. The only thing that could be different is that 
there could be some different measure of charge than electric but the dynamics of the force will however be determined by Maxwell equation. There is hence 
no sensible reason to have two different forces having the same dynamics. 
Thus electromagnetic force is the unique non-universal long range force. \\

Gravity and electromagnetic force are the only two long range forces. There 
could exist no other long range force and hence search for new 
force should be directed at short range. They are classical forces;i.e. they accord to classical field theory description. When we relax the property (b), 
force will be of short range. It would be a quantum force requiring quantum field theory description. How to make force short range, confined to a region ? The only way it could be done is by making either propagator massive or coupling (strength) of the force diminishing/enhancing with decreasing/increasing 
distance.   \\

\subsection{Weak Force}

Here we relax (b) and retain (a) partially, so we have universal linkage to all massive particles but short 
range. Linkage is though universal but only to all massive particles. As 
argued above linkage to massless particles could only be effected through 
curvature of spacetime which can not be kept confined to short range. So 
massless particles like photons and gravitons are excluded from interaction 
with short range force. The force should interact with all massive particles irrespective of any other parameter like charge it has or not. Such a force can 
then be the weak force~\cite{das}. Its 
propagator is massive and since it interacts with both electrically charged as well as neutral particles accordingly it also carries charge so as to 
conserve charge through the interaction. It is a non-abelian force which 
arises through symmetry breaking. There is a unified electro-weak force with 
gauge symmetry $SU(2)XU(1)$ which breaks into massive $SU(2)$ weak force 
and massless $U(1)$ electromagnetic force. Note however that $SU(2)$ is not a 
gauge symmetry. It is a broken symmetry which is responsible for making 
propagators massive. There is indeed a unification of electromagnetic and 
weak force. \\

In our perspective, there is complementarity between electric and weak forces 
because the former has a specific charge linkage but long range (massless propagator) while the latter has universal linkage to all massive particles but 
propagator is massive. It is therefore not surprising that the two have a 
unified description. However one would further expect that there may exist 
some duality between the two in some appropriate space.

\subsection{Strong Force}

Finally we relax both the properties (a) and (b);i.e. neither universal 
linkage nor long range. This is the force which links only to particles which 
have specific new charge and is also of short range. Let us resort to the 
other 
alternative of confinement where propagator could be massless but coupling 
diminishes/enhances exponentially for decreasing/increasing distance. As 
$r\to0$, force also tends to zero. This is what is called the asymptotic 
freedom~\cite{gross}. This can then be the Strong force. It is a non-abelian 
gauge field with a non-abelian new colour charge and true $SU(3)$ gauge 
symmetry. This is complete opposite and complementary of gravity. If we go by 
the previous example, there is a strong case for unification of gravity and strong force !   \\

This exhausts all the four possibilities representing the four basic forces.\\ 

\section{Suggestions and Predictions}

We have knit together all the four basic forces with gravity being the mother 
and the other three emerging out of it in a natural manner. We have begun with 
the general principle of universality and have then probed what happens as 
layers of universality are peeled off. In this process we envision a unified 
view of all the four forces which is consistent with the fundamental physics as we know today. We shall now discuss the perspective and insight 
this gives rise to.\\

Gravity and electromagnetic are the two unique long range forces. However, it 
is not possible to prove uniqueness of short range forces. What we have shown 
is that weak and strong forces do comply with our general principle but we 
can't prove that they are unique. The mother force is a spin 2 field and it can in 
general include spin 1 vector and spin 0 scalar field. Short range will ask 
for vector field to be non-abelian. Abelian field can not be confined because 
it carries no hooks in terms of indices which can be engaged to trap it. On 
non-abelian vector field, one can impose not only $SU(2)$ and $SU(3)$ but any 
kind of symmetry gauge or non-gauge. Hence the question, how many remains open 
until we are able to hit upon some principle which picks up these two or more.
In addition, we could of course have scalar fields like Higgs as predicted by 
the standard model of particle physics~\cite{das}. They will also have to  
be of short range.  \\

Thus if we are to look for new fifth force, search must not be directed at 
long range but to short range and it will have to be a non-abelian vector or 
a scalar field.\\

Whatever interacts with weak force must be massive and hence neutrino must 
have non-zero mass howsoever small. Similarly all massive particles must 
respond to it and thereby both left as well as right handed electrons must 
participate in the weak interaction. Though interaction with right handed 
electrons has not yet been observed however our paradigm does demand it and it is also required by supersymmetry~\cite{moh}. It is hoped that it may show up 
at higher energy yet to be achieved. \\

As noted above, there is complementarity between the forces, electric and weak and gravity 
and strong. Electric force is long range but not universal while weak 
is universal but short range. Similarly gravity is both universal and long 
range while strong is neither. Drawing upon the electro-weak unification, 
there is a strong suggestion that we should seek unification of gravity and 
strong force. For that we may have first to have a quantum 
description of gravity. That is the big question all by itself. The complementarity also points to some kind of duality between the forces in some appropriate space. It is interesting to note that this 
suggestion/conjecture finds good resonance in some of the established results. On the one hand we have electro-weak 
unification~\cite{das} and on the other AdS/CFT correspondence which envision 
gravity (AdS) living on the boundary of bulk spacetime harbouring strong 
force (Quantum Chromodynamics)~\cite{mal}. Further a common feature between gravity and strong force is also indicated by the presence of spin 2 particle in 
string theory description of quantum chromodynamics. It would therefore be not 
out of place to seek duality between the forces. \\

The overall picture that emerges is as follows. Imagine a diamond shaped 
figure with gravity (Linkage: completely universal to all massive and massless particles; Propagator: massless) sitting at the top. Then on the two 
sides are electric (Linkage: electric charge; Propagator: massless) and weak 
(Linkage: partially universal, only to massive particles; Propagator: massive) and at the bottom is strong (Linkage: colour charge; Propagator: massless with running coupling). Unification/duality is suggested for the forces facing each-other. One might however ask that it is 
an interesting way of looking at the forces but what 
does it really lead to ? To this end we would like to say that this viewpoint 
is not only insightful but also points, as alluded above, guiding finger to 
clear and definitive suggestions and predictions. They are as follows: \\

1. If we are to search for new force, it should be directed at short range 
and the new force will be non-abelian vector field or a scalar field. That 
is to probe uniqueness of weak and strong forces. \\

2. Neutrino must have non-zero mass howsoever small and right handed electrons 
must interact with the weak force. \\

3. There must exist duality relations between electric and weak, and between 
gravity and strong. This indicates that unification of forces may also follow 
this order which has already been achieved for electric and weak in electro-weak force.  \\

It would be interesting to pursue some of these leads. The ultimate aim is 
to achieve and to understand unification of all the forces. Except the mother 
force gravity, the other three accord to quantum description. In some sense a
programme of quantization of gravity should also include unification of 
forces. This is the spirit of string theory~\cite{john} while loop quantum 
gravity~\cite{ast} attempts to quantize Einstein's gravity. In the latter, 
unification is though not directly addressed yet it has to be a theory of 
quantum spacetime which should, we believe, encompass everything that exists 
in spacetime including all the forces. Similar sentiment echoes in a recent 
essay on gauge/gravity duality where it is argued that a 
non-Abelian gauge theory including weak and strong forces imbibes in some 
hidden manner a theory of quantum gravity~\cite{hor}. This 
perhaps means that a non-abelian gauge theory may at deep down have anchoring 
in quantum spacetime.\\ 

At any rate, our simple guiding principle of universality and its subsequent 
peeling off presents a tantalizingly simple and insightful paradigm. It is not only compatible with the fundamental physics as we currently understand but 
also makes remarkably clear and interesting suggestions/predictions.   \\

\noindent{\bf Acknowledgment}: This perspective has evolved over a period of 
time and 
during which the author had the benefit of bouncing these ideas with many 
colleagues which included M. Sami, Parampreet Singh, S. R. Chodhary, Jnan 
Maharana, Bindu Bambah, Chandrasekhar Mukku, P. P. Divakaran, Shyam Date and 
T. Padmanabhan. I thank them all. Special thanks are due to Shaym Date for 
reading the manuscript and making many clarifying comments which have 
resulted in improved clarity.

\end{document}